\documentclass[journal=jacs,manuscript=article]{achemso}
\usepackage[T1]{fontenc}
\usepackage{amsmath}

\usepackage{color}

\usepackage{times}
\usepackage[version=3]{mhchem} 
\usepackage[T1]{fontenc}       

\author{Semion K. Saikin}
\affiliation{Department of Chemistry and Chemical Biology, Harvard University, Cambridge, MA 02138, USA}
\email{ssaykin@fas.harvard.edu}
\author{Mars A. Shakirov}
\affiliation{Institute of Physics, Kazan Federal University, Kazan, 420008, Russian Federation}
\author{Christoph Kreisbeck}
\affiliation{Department of Chemistry and Chemical Biology, Harvard University, Cambridge, MA 02138, USA}
\author{Uri Peskin}
\affiliation{Schulich Faculty of Chemistry, Technion-Israel Institute of Technology, Haifa 32000, Israel}
\author{Yurii N. Proshin}
\affiliation{Institute of Physics, Kazan Federal University, Kazan, 420008, Russian Federation}
\author{Al\'an Aspuru-Guzik}
\affiliation{Department of Chemistry and Chemical Biology, Harvard University, Cambridge, MA 02138, USA}
\altaffiliation{Canadian Institute for Advanced Research (CIFAR) Senior Fellow}

\email{aspuru@chemistry.harvard.edu}

\title{On the Long-Range Exciton Transport in H-Aggregated Heterotriangulene Chains}


\begin{document}

\begin{abstract}
  Self-assembled aggregates of pigment molecules are potential building blocks for excitonic circuits that find their application in energy conversion and optical signal processing. Recent experimental studies of one-dimensional heterotriangulene supramolecular aggregates suggested that singlet excitons in these structures can propagate on several micron distances. We explore this possibility theoretically by combining electronic structure calculations with microscopic models for exciton transport. Detailed characterization of the structural disorder and exciton decoherence is provided. We argue that conventional exciton transport models give about an order of magnitude shorter estimates for the exciton propagation length which suggest that there are other possible explanations of the experimental results.
\end{abstract}

\section{Introduction}

Molecular aggregates are a broad class of structures, where the constituent molecules are bound by non-covalent interactions. A canonical example of just such structures are molecular crystals with almost perfect 3D packing structures.\cite{MolCrystl_Book} In addition to that, many molecules combine in a variety of lower dimensional structures such as 2D layers,\cite{PaGaHi08_5946_,Wurthner2011} tubular aggregates,\cite{KiDa06_20363_,Caram2016} and 1D chains.\cite{MaLiWu11_648_,Kivala2013} Optical properties of molecular aggregates composed of organic dyes depend dramatically on the form of aggregation. For instance, molecular absorption peaks shift to the long or to the short wavelengths relative to the position of the monomer absorption forming so-called J- and H-aggregates respectively. \cite{Jelley1936, Scheibe1936} In a more general case, multiple absorption peaks are formed. \cite{David_ACSEnLett2016}  These spectral changes, occurring due to the Coulomb interaction between molecular excitations, reflect modifications in the excited states properties due to the aggregation and can be used in the design of photon processing devices and excitonic circuits.\cite{SKS_NanoPh2013}

Indeed, properties of  aggregated organic dyes, which have electronic excitations within the optical band, can be tuned through a broad range of absorption frequencies by a proper choice of pigment molecules.\cite{Walker_NanoLett,Fabian201036} The reported exciton energy transport on micron length scales\cite{Yossi_PCCP2014,Haedler2015,Caram2016} opens an opportunity for using these structures in devices. For example, one can think of converting photons to excitons, process them, and finally return the signal back to the photon mode. As compared to photons, excitons interact with each other,\cite{Akselrod2010} which potentially allows us for a nonlinear signal control. As another example, one can think of creating structures with topologically protected bands, where excitons propagate in a particular direction and are not sensitive to structural defects.\cite{Yuen-Zhou2014, Joel_NatComm2016} Moreover, exciton propagation can be confined much below the diffraction limit to the extent of single molecules. This is, actually, a case of photosynthetic organisms that use aggregates of pigments as light-harvesting antenna.\cite{Scholes2011,Huh_JACS,Sawaya2015} In these organisms, the absorbed energy of light is transferred through large excitonic assemblies with high precision, and the transfer process is very robust to different types of structural disorder.

In Ref \cite{Haedler2015} the authors report that singlet excitons in heterotriangulene chains can propagate on several micron distances at room temperatures. They suggest that in order to achieve this, the excitons should move coherently on a sufficient part of the propagation length. An interesting property of this system is that due to the $\pi-\pi$ stacking, the dye molecules form an H-aggregate.

While the most of recent research of artificial molecular aggregates was focused on J-aggregates due to their distinct optical signatures such as a strong absorption and fluorescence, line narrowing, and reduction of the Stokes shift, their counterparts -- H-aggregates -- also possess intriguing characteristics. The absorption peak in H-aggregates is shifted to the top of the exciton band leaving the low-lying exciton levels optically inactive or "dark." Therefore, bright excitons created after a photon absorption will relax into a set of dark states and will remain there for sufficiently longer times, as compared to J-aggregates, due to the suppression of the photoluminescence decay channel. In contrast to the triplet excitons, where intermolecular coupling decays exponentially with the distance, the exciton coupling in H-aggregates decays polynomially. In addition, the dark singlet excitons are delocalized over multiple molecules,\cite{David_ACSEnLett2016} which can also affect the transport properties.

In this article we analyze dynamics of electronic excitations in 1D molecular chains using a microscopic model of triangulene aggregates with structural disorder. Then, the exciton propagation distance is computed using two complementary models: the transport model by Haken, Reneiker, and Stroble (HRS)\cite{Haken1972,Haken1973}; and a more sophisticated Hierarchical Equations of Motion (HEOM) approach.\cite{tanimura1989a,tanimura2012a,hu2011a,shi2009a,IsFl09_234111_} We find that in order to obtain a micron scale propagation length the interaction of the excitons with the environment and the structural disorder should be strongly suppressed. Finally, we estimate the range of structural and dynamic disorder that would satisfy the requirement of exciton transport on a micron distance.

\section{Results}

\subsection{Model}

\begin{figure}
\begin{center}
\includegraphics[scale=1.5]{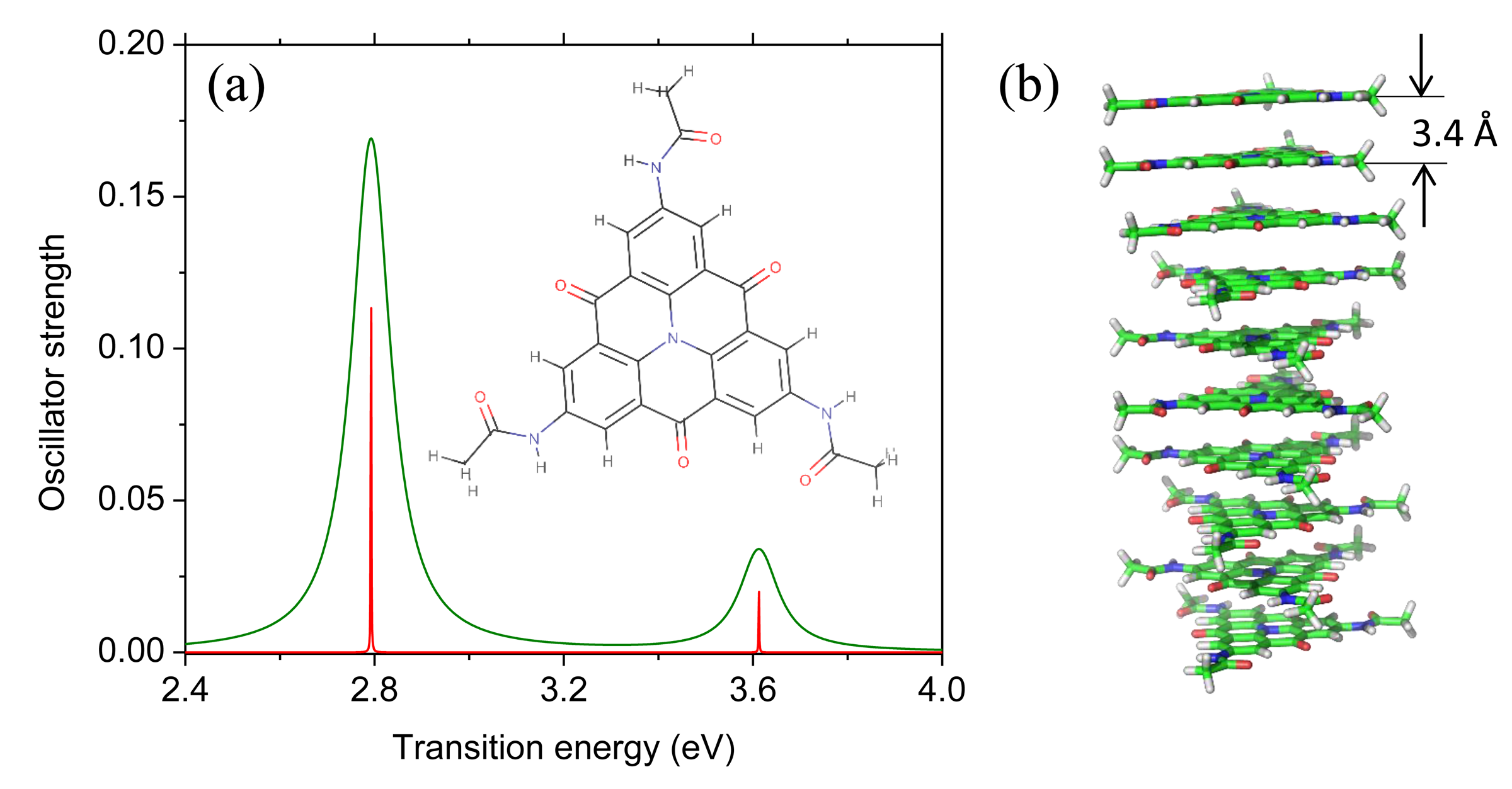}
\caption{(a) Electronic excitation spectrum of an isolated carbonyl-bridged triarylamine (CBT) dye molecule computed with TDDFT. The vibronic structure and the Jahn-Teller distortion effect are not included. The red peaks correspond to the computed transitions, while the green envelop line shows a profile smoothed Lorentzian. Inset: a chemical structure of CBT. (b) An illustration of 1D self-assembled chain of CBT dyes. The intermolecular distance corresponds to a $\pi - \pi$ stacking. }
\label{fig:struct}
\end{center}
\end{figure}

Our model system is a chain of carbonyl-bridged triarylamine (CBT) dyes, Fig.~\ref{fig:struct}. These molecules belong to a broader class of triangulenes \cite{Clar1953}, where the carbonyl bridges stabilize the planar structure of the molecule. In comparison to unsubstituted triangulenes \cite{Pavlicek2017} the ground electronic state of CBT is a singlet. It has been shown that CBT molecules form stacks due to the interaction between the $\pi$-orbitals of different molecules ($\pi-\pi$ stacking). \cite{Kivala2013} A proper choice of the side chains attached at the vertices of the triangle controls the displacement as well as the rotation\cite{Haedler_JACS2016} of the molecules relative to each other. Moreover the same sidechains chosen as dyes can be use as reporters for studies of exciton dynamics in the chains. For instance, the authors of ref.~ \cite{Haedler2015} used 4-(5-hexyl-2,2'-bithiophene)naphtalimide (NIBT) dyes as side chains and as fluorescence reporters.

\textit{Electronic excitation spectra.}
The electronic excitation spectra of single CBT dyes were computed using time-dependent density functional theory (TDDFT) as it implemented in Turbomole~6.0.\cite{TM_general} We used triple-$\zeta$ def2-TZVP basis set\cite{BS_def2} and the B3LYP hybrid functional\cite{B3LYP} for the ground structure optimization and computation of the response. \footnote{Using a PBE0 functional results in a systematic shift of electronic excitations by about 100meV.} Among 10~lowest electronic excitations, only two strong doubly degenerate transition in the visible - near UV part of the spectrum are found (Figure~\ref{fig:struct}(b)).

The lowest vertical electronic excitation in CBT dyes is optically active and double degenerate. The associated transition dipole is $\mu_{\rm TDDFT}=2.8$~Debye (D), and the transition frequency is 2.79~eV (Fig.~\ref{fig:struct}(b)). This transition is composed of HOMO $\rightarrow$ LUMO by 97\%. Rotations of the side chains about the C-N bonds result in the splitting of the lowest state by approximately 100~meV (see SI1 for details). The next strong optically active transition is at 3.6~eV, which leaves about 0.8~eV transparency window above the lowest excitation.

\textit{Intermolecular coupling.} The proximity of the CBT molecules in a chain modifies their electronic excitation spectra. There are two main contributions to the interaction between the electronic excitations, the F\"orster or near-field interaction\cite{Foe65_93_} and the Dexter interaction \cite{Dexter_JCP1953} that involves electron exchange between the molecules. The former coupling dominates the singlet-singlet interaction in aggregates of pigments. Both interactions originate in the Coulomb coupling between electronic states of the molecules and can be written in a two-electron approximation as
\begin{align}
    V_{\rm F}&= \int \mathrm{d}^3x \int \mathrm{d}^3y \, \rho^{(1)}_{ge}(x,x) \frac{e^2}{|x-y|}\rho^{(2)}_{eg}(y,y), \label{V_F}\\
    V_{\rm D}&= \int \mathrm{d}^3x \int \mathrm{d}^3y \, \rho^{(1)}_{ge}(x,y) \frac{e^2}{|x-y|}\rho^{(2)}_{eg}(y,x),\label{V_D}
\end{align}
\noindent where $\rho^{(m)}_{ge}(x,y)$  is an electron transition density between a ground, $g$, and excited, $e$, state of a molecule, $m$. The coordinates $x$ and $y$ correspond to two electrons.

\textit{Extended tripoles.} Microscopic calculations of excitonic couplings in a large aggregate of pigments, using the outlined above procedure, are computationally intensive. Therefore, additional approximations using phenomenological models are frequently applied with the parameters fitted to the microscopic calculations.  If the distance between the molecules is sufficiently larger than the extent of the intramolecular excitation, a point dipole approximation provides reasonable results. For shorter distances, various versions of monopole expansions are used. For instance, for cyanine dyes molecules it is an extended dipole model.\cite{Ko96__} For long oligomers, a dipole chain model was applied.\cite{Manas1998} These models can be linked to early works of Platt for cata-condensed aromatic molecules.\cite{Platt1949}

\begin{figure}
\begin{center}
\includegraphics[scale=2]{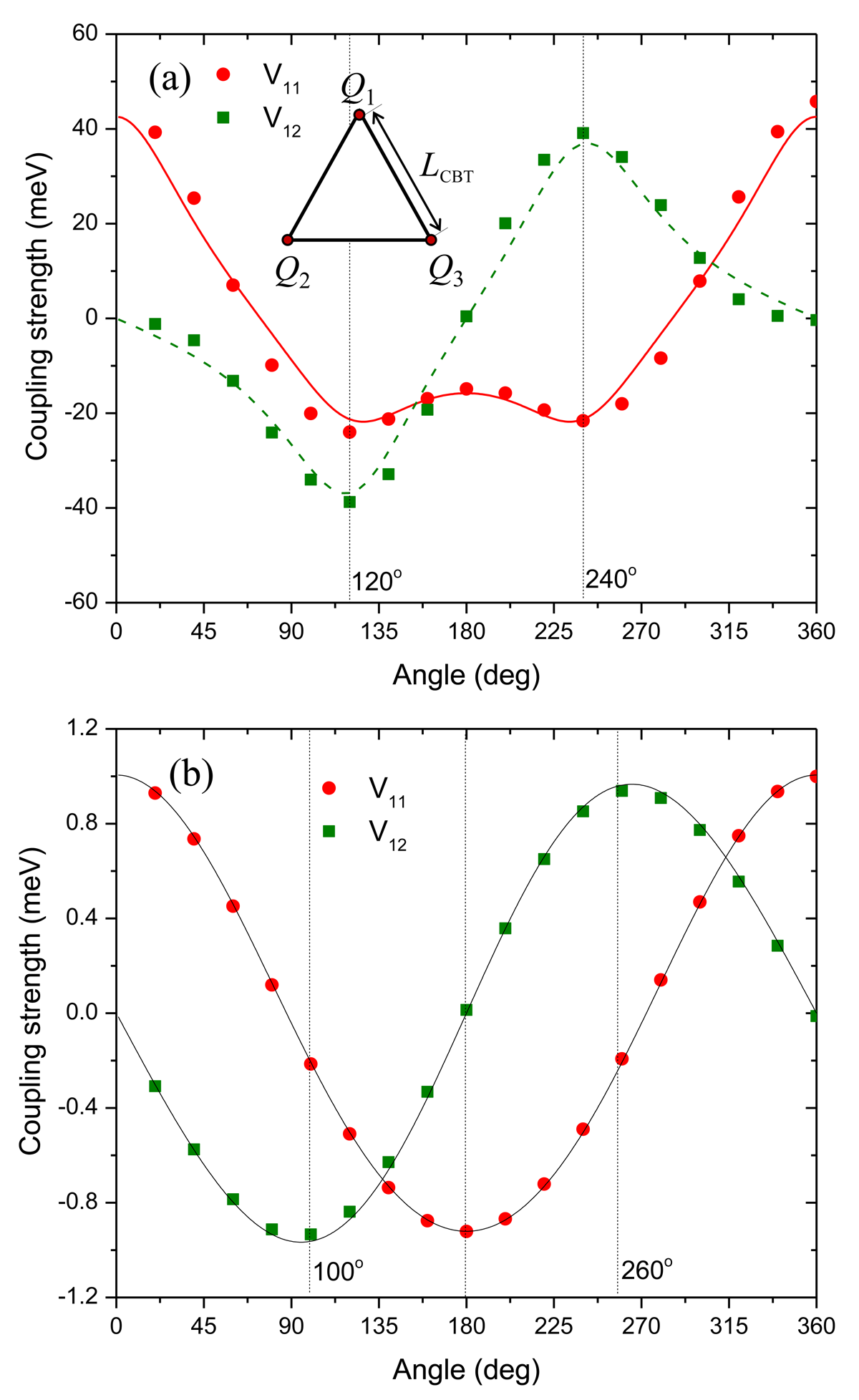}
\caption{Fitting of the coupling in a CBT molecular dimer computed with the molecular transition densities to the extended tripole model (inset). The figures show the couplings between the same tripoles, $V_{11}$, and the different tripoles, $V_{12}$, as a function of the relative angle for two different distances, $d$. (a) Nearest neighbors, $d=3.4$~\AA. The coupling reflects the trigonal symmetry of the molecules. (b) The distance $d=17$~\AA. The coupling is similar to the interaction between extended transition dipoles.}
\label{fig:inter_dimer}
\end{center}
\end{figure}

For CBT molecules we introduced an extended tripole model. This is the simplest approximation that accounts for the trigonal symmetry of the molecule. In the model, each excitation is characterized by three transition charges located in vertices of an effective triangle, Fig.~\ref{fig:inter_dimer}(inset). The transition charges of the tripoles should satisfy: (a) the total charge neutrality, (b) a convergence to point dipoles at long distances, and (c) the state orthogonality. These conditions determine the relations between the two sets of charges as $t_1=(2q/\sqrt{3},-q/\sqrt{3},-q/\sqrt{3})$  and $t_2=(0,q,-q)$, where $q=\mu/l$, $\mu$  is a transition dipole computed for a far-field limit, and $l$ is an effective distance between the charges that should be fitted.

\textit{Fitting procedure.} We fit the extended tripole model to the intermolecular coupling in a dimer of CBT dyes displaced along the C$_{3}$ symmetry axis on the distance range $2.5-17$~\AA. The analytical formulas for two interacting tripoles displaced along z-axis by distance $d$ and rotated relative to each other by angle $\phi$ are
\begin{equation}
    V_{11} = \frac{2 \kappa q^2 }{\sqrt{(2/3)l^2(1-\cos\phi)+z^2}}-\frac{\kappa q^2}{\sqrt{(2/3)l^2(1-\cos \varphi)+z^2}} - \frac{\kappa q^2}{\sqrt{(2/3)l^2(1-\cos \theta)+z^2}}, \label{V_11}
\end{equation}
and
\begin{equation}
    V_{12} = \frac{\sqrt{3}\kappa q^2}{\sqrt{(2/3)l^2(1-\cos \varphi) +z^2}} - \frac{\sqrt{3}\kappa q^2}{\sqrt{(2/3)l^2(1-\cos \theta)+z^2}},\label{V_12}
\end{equation}
\noindent where $V_{11}$ and $V_{12}$ correspond to the interactions between the same and the different transitions respectively, $\kappa=14.38$~V$\cdot$~\AA/e is a coupling coefficient, and two complementary angles are defined as $\varphi = 2\pi/3+\phi$, $\theta = 2\pi/3-\phi$. Figure~\ref{fig:inter_dimer} shows the fit of Eqs.~\ref{V_11},\ref{V_12} to the couplings computed with the molecular transition densities for two values of the intermolecular distance. While the short distance interaction shows $C_{3}$ symmetry, it tends to a $C_2$ symmetry of dipolar interaction, when the distance is increased. In the fitting model, the distance between the transition charges equals $L_{\rm CBT}=7.3$~\AA\ (see inset in Fig.~\ref{fig:inter_dimer}) and the value of the transition charge is $q=0.08$~e. We also rescaled the intermolecular distance as $\tilde{z}  = z - \delta z$ in order to account for a finite extent of the electron density in the $z$-direction. The value of $\delta z = 0.4$~\AA.

\subsection{Hamiltonian and Spectra}

A two-bands electronic Hamiltonian is constructed in a tight-binding form as
\begin{equation}
    H = \sum_{l,\alpha}\omega_{l\alpha}|l\alpha\rangle\langle l\alpha| + \sum_{l, m, \alpha \neq \beta} k_{lm;\alpha\beta}(|l\alpha\rangle\langle m\beta| + |m \beta\rangle\langle l\alpha|),\label{Ham_2band}
\end{equation}
\noindent where we consider a single exciton manifold only, ($l$, $m$) are indices of the molecules, and ($\alpha$, $\beta$) correspond to the two excited states of each molecule. In the basis of single molecules, the states are $|l\alpha\rangle = |g_1 g_2 ...e_{l,\alpha}g_{l+1}...\rangle $, $\omega_{l,\alpha}$   is a frequency of single intramolecular transitions or site energies, and $k_{lm;\alpha\beta}$ is a F\"orster coupling between the corresponding transitions $l$ and $m$.  In this approximation the two transitions on a single CBT molecule are electronically decoupled and the vibronic interaction is not included.

The presence of electronic coupling between the two excitations on a single CBT molecule can lead to intra-chromophore energy transfer, even when the intra-molecular vibronic coupling is negligible. Dipoles in the molecular surroundings may respond differently to different excited tripole states on the same molecule, thus driving on-sites intramolecular transitions. However, in the typical scenario, where the environment reorganization energy associated with these transitions is large with respect to the intramolecular electronic coupling energy, the respective intramolecular transitions should have no apparent effect on the exciton propagation between different molecules through the aggregate. In this case, one can disregard the on-site electronic coupling between the states $\alpha$  and $\beta$, as implemented in the Hamiltonian (5). Things would be different, however, if the electronic coupling energy between (intra-molecular and inter-molecular) transitions would be large with respect to the environment reorganization energy. In this limit, exciton propagation rates through the aggregate could depend strongly on the on-site electronic coupling. This regime is beyond the scope of the present model (see SI2 for more details).

The exciton states of the aggregate are the eigenstates of the Hamiltonian~(\ref{Ham_2band}). The interaction with optical far fields is considered in a point dipole approximation. The exciton transition dipoles are defined in the site basis as $\mathbf{\mu}_X = S \mathbf{\mu} $, where $S$ are eigenvectors of the Hamiltonian~(\ref{Ham_2band}) and $\mathbf{\mu}$ is a vector of molecular transition dipoles \cite{Valleau_ACS2014}. Then, the strength of electronic excitations is proportional to the square of the corresponding exciton transition dipole $I\propto |\mathbf{\mu}_X|^2$. In the result, the electronic excitation spectra of the aggregates are computed as a convolution of the excitonic transitions with a Lorentzian lineshape.

\begin{figure}
\begin{center}
\includegraphics[scale=1.5]{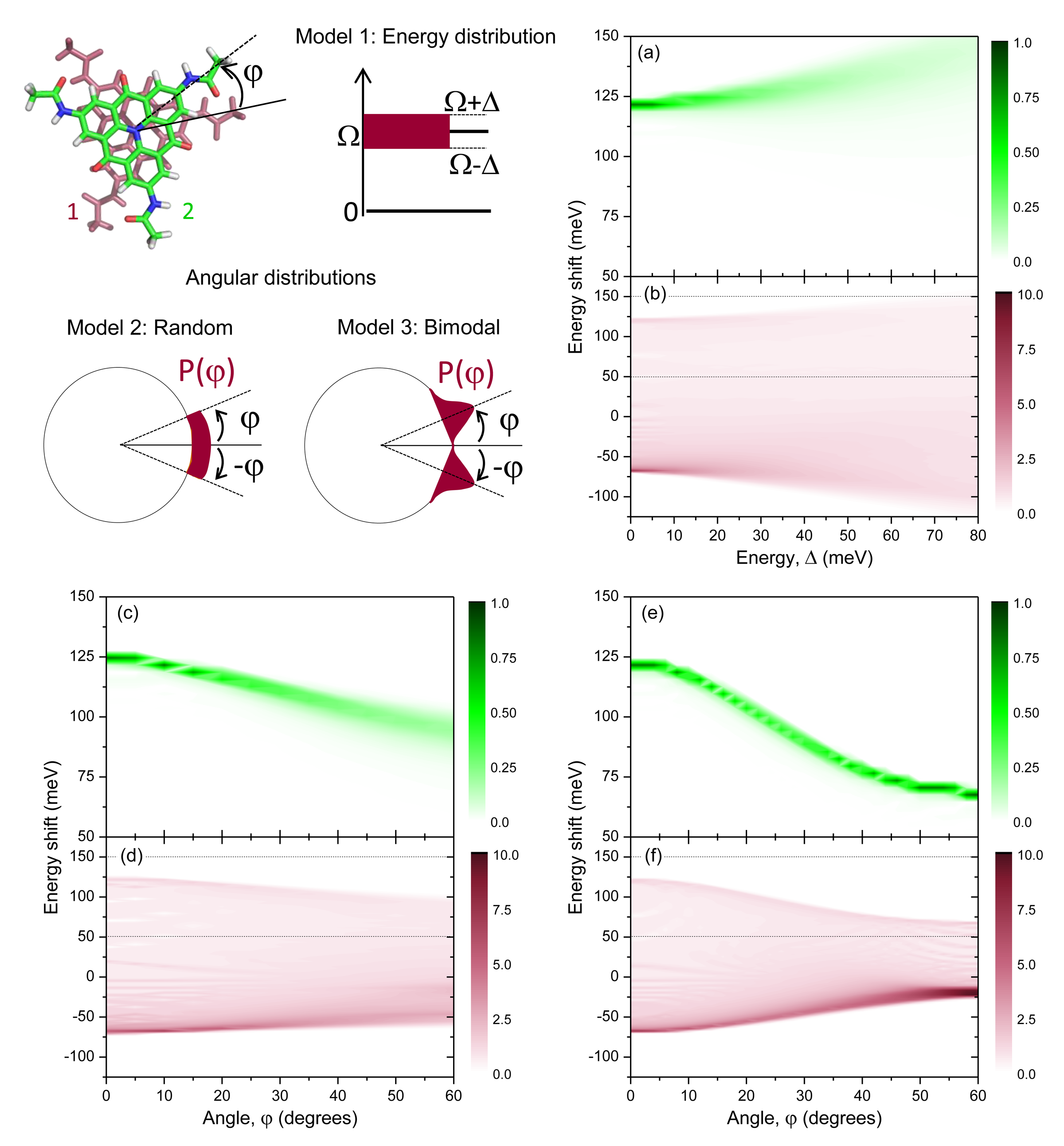}
\caption{Structural disorder models and exciton band structure of CBT chains. Electronic excitation spectra (a,c,f) and exciton densities of states (b,d,e) for different types of disorder in  CBT chains composed of 100 molecules. The spectra are averaged over 1000 disorder realizations. (a,b) site energy distribution, (c,d) random angular distribution, (e,f) bimodal angular distribution. The vertical scale in the figures correspond to the shift of the exciton energy relative to the lowest electronic excitation in single molecules. The color scales: the peak area of the excitation spectra is normalized to 1, the DOS scale corresponds to the fraction of exciton states in \% per 1 meV. The spectral window of the electronic excitation spectra shown in the DOS profiles by dashed lines.}
\label{fig:spectra}
\end{center}
\end{figure}

\textit{Disorder.} While in a perfect 1D chain of molecules the site energies should be identical, variations in the local molecular environment introduce static fluctuations of these values.  Similarly, the intermolecular coupling constants are sensitive to the deviation of the structure from a perfect periodicity.  Here, we consider two types of structural disorders: (a) static fluctuations of site energies due to local imperfections and (b) variations in the relative orientations of the pigment molecules. These disorder models are schematically illustrated in Fig.~\ref{fig:spectra}. While we fix the intramolecular distance at $3.4$ \AA\ and do not allow for relative tilts and displacements of the CBT molecules, the molecules can be rotated around the C$_3$ axis. The type of the side chains can provide additional constraints on this degree of freedom. For instance, bulky chains will tend to repulse, which would forbid a perfect alignment of the molecules on top of each other. However, if the side chains are hydrophobic, they would tend to cluster just to minimize the surface area. Moreover, the side chains can couple to each other, for instance through hydrogen bonds such as NIBT molecules in Ref.~\cite{Haedler2015}. As representative models we use two types of angular distribution: a random distribution, where the angular distribution probability is a constant, and a bimodal distribution, where the distribution function is composed of two Gaussians of a fixed width.

\textit{Spectra.} Figure~\ref{fig:spectra}(a-f) shows how the electronic excitations spectra and densities of states (DOS) in disordered CBT chains depend on different types of disorder. The distribution of the site energies broadens the excitonic band and homogenizes the DOS, Fig.~\ref{fig:spectra}(a-b). The bright exciton states remain closer to the upper edge. However, in analogy with J-aggregates the top most states become dark with increasing the disorder. The rotational disorders reduce the hypsochromic shift of the bright excitonic subband and also broaden it, Fig.~\ref{fig:spectra}(c-f). This subband corresponds to the upper edge of the total exciton band, as can be seen from the DOS profiles, while the maximal density of exciton states is at the lower edge. The DOS profiles shrink with the increase of the angular disorder. In the bimodal distribution, the shift of the bright subband and the modulation of DOS reflect the changes in the nearest-neighbor interaction for the most probable angle as the width of the distribution is fixed at 5 degrees. For all studied cases of disorder the structure remains an H-aggregate with lowest electronic excitations completely dark. The variation of DOS edges is related to the strength of intermolecular coupling, and its value varies within 50~meV.

\subsection{Exciton Transport}
\textit{Dephasing.} Interactions with intramolecular and intermolecular vibrations modulate both the site energies and the coupling terms and introduce exciton decoherence and relaxation channels into the system.
There are a number of theoretical methods that handle the dissipative dynamics in just such a system \cite{Haken1972,IsFl09_234111_,RoStEi10_5060_,ScTe91_421_,SKS_NanoPh2013}. For the interaction of excitons with dynamical environments we consider two complementary models. In the HRS model \cite{Haken1972,Haken1973} the coupling of the exciton to the vibrational modes is considered as a white noise which results in the fluctuations of the site energies. This model involves an infinite temperature approximation, where the exciton energy spreads over the band equally in both directions at long times. For a given dephasing rate, this model provides us with an upper bound estimates for the transport coefficients. The second model, HEOM, describes non-Markovian exciton dynamics due to the electron-vibrational coupling using equations of motion. It also accounts for a proper thermalization of the system, accurately takes into account the finite time-scale of the reorganization process, and works for a wide parameter range for the coupling to the environment.

\textit{HRS model.} We implemented the HRS model using a Monte Carlo propagation\cite{Molmer1993,Stephanie_JCP12} of a stochastic Schr\"odinger equation\cite{kampen2007} with a phase noise
\begin{equation}
    |\dot{\psi}(t)\rangle = (-i \tilde{H} + \sum_{m \alpha}f_{m \alpha}(t)\hat{C}_{m \alpha}) |\psi(t)\rangle,
    \label{Schrod}
\end{equation}
where $\hat{C}_{m \alpha} = -\sqrt{\Gamma}|m\alpha\rangle\langle m\alpha|$ is a noise operator acting on sites in the molecular basis, $f_{m \alpha}(t)$ is a normalized stochastic fluctuation term, $\Gamma$ is a dephasing rate, and the Hamiltonian $\tilde{H} = H - i/2\sum_{m\alpha}\hat{C}_{m \alpha}^\dagger\hat{C}_{m \alpha}$ includes a non-Hermitian term due to fluctuations \cite{Molmer1993}. Then, the stochastic wavefunction $|\psi(t)\rangle$ is computed for different realizations of noise and structural disorder. Finally, the exciton population dynamics is characterized by the second moment of the exciton distribution function computed as $M_2(t) = \sum_{m,\alpha} (m-m_0)^2 |m\alpha\rangle\langle m\alpha|$. In this model, an exciton initially localized on the site $(m_0,\alpha)$ propagates ballistically. Then, on timescales of exciton dephasing its motion becomes diffusive. The time dynamics of the second moment, $M_2(t)$, of the exciton function characterizes both ballistic and diffusive transport regimes. The second moment grows quadratically with time for ballistic excitons. Then, it transforms to a linear dependence for the diffusive motion.\cite{Reineker_ZfP1973,Stephanie_JCP12} As compared to an analytically solvable case,\cite{Reineker_ZfP1973} our model also describes structurally disordered systems.

In the case if only the dephasing model is included, one can describe the second moment using the following analytical expression
\begin{equation}
    M_2(t) = At + B(e^{-\kappa t}-1),
    \label{M2_eqn}
\end{equation}
where the coefficient $A$ characterizes the diffusive transport (for 1D transport the diffusion coefficient is $D=A/2$), $B$ represents a square of the ballistic exciton delocalization length, and $\kappa$ is a characteristic dephasing time. If the dephasing in the structure is due to the site energy fluctuations only, the linear in time term in Eq.~\ref{M2_eqn} should be cancelled at short times $t<<\kappa^{-1}$ . This imposes an additional condition on the coefficients. For long times, the second term in Eq.~\ref{M2_eqn} vanishes leaving only the diffusion component. In order to estimate exciton diffusion coefficients and coherent delocalization length from a finite time exciton dynamics we assumed that this equation is also applicable for disordered structures. It should be noted that we consider only cases where the energy fluctuations associated with the structure disorder are less or comparable to the energies of dephasing fluctuations. If the static energy fluctuations become large the exciton transport become subdiffusive that cannot be described by Eq.~\ref{M2_eqn}.

We computed exciton dynamics in the CBT chains for all 3 models of structural disorder shown in Fig.~\ref{fig:spectra}. In our approach, electronic excitations initially localized on a single molecule in the center of a 101 molecule chain were propagated using Eq.~\ref{Schrod} for 300 femtoseconds. The propagation time step was equal to 1 attosecond in order to conserve the norm of the wavefunction. The wavefunction dephasing rate was scanned over the range $[10-100]$~meV which corresponds to $6.5-65$~fs range for the exciton decoherence time. We verified that for all sets of the model parameters the exciton transport is either in a ballistic or a diffusive regime, and the effects of localization due to the structural disorder are negligible. We also checked that on the propagation timescale the effects of the finite length of the chain are small.

\begin{figure}
\begin{center}
\includegraphics[scale=1.5]{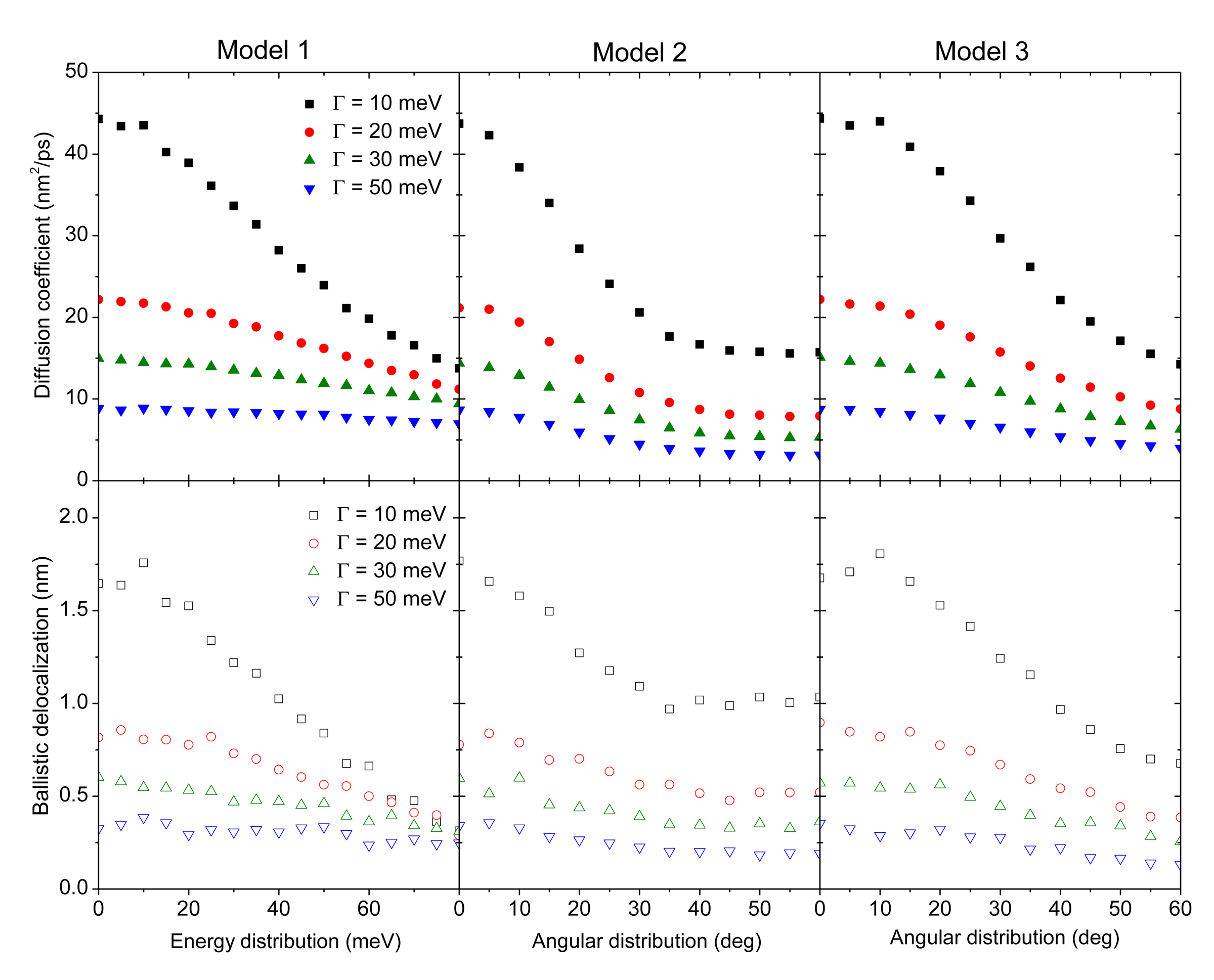}
\caption{Computed exciton diffusion coefficients and ballistic delocalization lengths as a function of disorder parameters. Exciton propagation has been averaged over 1000 trajectories with different realizations of disorder. The disorder models are shown in Fig.~\ref{fig:spectra}, and $\Gamma$ is a rate of dynamical fluctuations.}
\label{fig:diffusion}
\end{center}
\end{figure}

In Fig.~\ref{fig:diffusion} we show the computed diffusion coefficients and the initial exciton delocalization lengths for different values of structural disorder and dynamical dephasing. The range of the computed diffusion coefficients is within $3-45$~nm$^2$/ps. This would result in exciton propagation over $80-300$~nm within 1 nanosecond. For all disorder models the structural variations slow down the exciton diffusion by about 50-60\% as compared to the ordered case. The initial delocalization length reaches about 1.7 nm which corresponds to 6 molecules in the chain. This parameter is also sensitive to the dynamical and structural disorder.

In order to investigate the influence of processes that are neglected in the pure dephasing picture, such as thermalization, we analyze the diffusive dynamics in a CBT chain within the HEOM technique. HEOM is a non-perturbative approach for simulating the excitation energy transfer dynamics that properly describes energy relaxation and takes into account the finite time scale of the reorganization process during which the vibrational coordinates relax to its equilibrium position in the excited state potential energy surface. Due to the computational complexity of HEOM, which scales exponentially with increasing system size and truncation level of the hierarchy, we restrict the subsequent analysis to a chain of 31~pigments and benchmark the HRS model for the non-disordered chain only. The calculations are done using the \textit{QMaster} software-package \cite{kreisbeck2014a}, which provides a high-performance implementation of HEOM \cite{kreisbeck2011a,kreisbeck2012a,kreisbeck2013a}. We tested convergence of the hierarchy depth by comparing different truncation levels. We exclude potential artifacts induced by the relatively small size of the considered chain by comparing to a longer chain comprising of 51 pigments. Due to the aforementioned computational restrictions the latter benchmark was done at lower truncation level for which the hierarchy is on the edge of convergence. We characterize the system-bath interaction using a coarse grained single peak shifted-Drude-Lorentz spectral density
\begin{equation}
 J(\omega)=\sum_{s=\pm}\frac{\gamma\lambda\omega}{\gamma^2+(\omega+s\Omega)^2},
\end{equation}
where we adjust the reorganization time-scale~$\gamma^{-1}=20$~fs and the center of the shifted peak~$\Omega=920$~cm$^{-1}$ such that we reproduce the pure-dephasing rate (assuming $T=300$~K) $\Gamma=10$~meV used in the HRS model. Motivated by the Stokes shift of 1337~cm$^{-1}$,\cite{Kivala2013} the reorganization energy is set to $\lambda=668.5$~cm$^{-1}$. Similar to the procedure of the HRS model, we compute the exciton dynamics for 300 femtoseconds and extract the diffusion coefficient $D=15.3$~nm$^2$/ps and ballistic exciton delocalization length is equal to $1.2$~nm (approx. 3 molecules in the chain) by fitting the population dynamics to Eq.~(7). Thus, the exciton is estimated to propagate over a distance of about 123~nm within 1~nanosecond, which is a bit less than half of the predicted distance of the HRS model.

\section{Discussion}
Our calculations for the intermolecular coupling strength and the shift of the absorption peak in CBT chains agree with the ones estimated in Ref.~\cite{Haedler2015}. However, the exciton diffusion coefficient computed within our model is not sufficient to describe exciton propagation over 3-4 microns within $2.4$ nanoseconds. If we assume only diffusive transport, the diffusion coefficient should be about an order of magnitude larger than our upper estimate. Moreover, the initial delocalization length of excitons is also sufficiently small. However, there are several model parameters that should be discussed in more details.

\textit{Role of side chains.} The NIBT side chains have two effects on the exciton transport in the chains. Firstly, they modify the spectra of CBT and introduces a relaxation channel. Secondly, they link the molecules through a network of hydrogen bonds between the carbonyl and the amine groups. In the result, the rotational disorder can be sufficiently reduced.

In order to estimate the contribution of the side chains in the electronic spectra of the composed CBT-(NIBT)$_3$ pigment, we have optimized the structure of the molecule and computed 40 lowest electronic states. We found that the excitation spectrum of a composed CBT-(NIBT)$_3$ molecule can be viewed as a superposition of the spectra of isolated molecules with some small shifts of the order of 100~meV due to the off-resonance coupling. The oscillator strengths of the intra-CBT transitions in a composed pigment are similar to the ones in an isolated molecule. The spectra of isolated CBT and NIBT molecules as compared to the spectrum of CBT-(NIBT)$_3$ are provided in SI1.

The hydrogen bonding between the pigments reduces the rotational disorder. However, this does not preclude from the 180 degrees rotation of the side chains. In this case, our model 3 in Fig.~\ref{fig:spectra} should be applicable. We estimate the relative rotation of the neighboring molecules by about 28 degrees, and we assume that the distribution of the angles within each peak is within 5 degrees. In this case the computed diffusion coefficient should be $25$~nm$^2$/ps only, see right panel in Fig.~\ref{fig:diffusion}.

\textit{Dephasing rate.} The dephasing rate is a free parameter in the HRS model. In general, this can be obtained from nonlinear pump-probe spectroscopy.\cite{} However, to our best knowledge there are no available nonlinear spectral data on CBT chains yet. Our estimates from the reorganization energy do not account for the changes due to the molecular packing which can reduce the dephasing rate. Therefore, it is useful to estimate the value assuming that only the site fluctuations dissipate the exciton coherent dynamics and the CBT chain has a perfect periodic structure (no structural disorder). In this case the ballistic and the diffusion coefficients are related to each other, and one can have analytical expressions for these coefficients.\cite{Reineker_ZfP1973} In order to have the second moment $M_2(t)=9$~$\mu$m$^2$ at 2.4~ns propagation time we need to have $\Gamma =0.25$~meV.

\textit{Triplet exciton states.} To estimate whether the intersystem crossing can play a role in the exciton transport in the chains, we computed energies of triplets in an isolated CBT molecule. The energy of the lowest state is about $2.2$eV which is sufficiently lower than the energy of the lowest singlet state in NIBT. Therefore, even if the singlet excitons in CBT are converted to higher lying triplets they should relax to the lowest state and do not transfer to NIBT singlets.

\textit{Transport through NIBT subsystem.} NIBT dyes can provide a substantial contribution to the exciton transport. The transition dipole of an isolated NIBT molecule computed with TDDFT is about 2 times larger than the dipole of CBT. However, the intermolecular distance is at least twice larger. If we assume a dipole-dipole coupling between the dyes, the coupling between two neighboring NIBT molecules should be approximately a half of the coupling in a CBT dimer. Then, taking to the account that there are three independent channels composed of NIBT, the energy transfer rate should be comparable to the transfer rate through the CBT system. However, the interaction between NIBT molecules is more sensitive to the disorder. Therefore, we expect that this value will be reduced several times.

\textit{Polariton modes.} Finally, we believe that the interaction of the exciton states with photon modes in this system should be explored in more details. It is possible that the exciton states hybridize with the photon forming attenuated polariton modes. In this case, even a small admixture of photons would reduce the dephasing rate of the excitations.

\section{Conclusions}
Recent experimental studies suggest that singlet excitons in supramolecular aggregates of pigments can propagate on several micron distances at room temperatures. Here, we provided a detailed theoretical analysis of exciton transport in 1D chains of triangulene dyes, which one of a few systems, where the long-range exciton transport has been reported. We apply two complementary transport models: Haken, Reineker and Stroble (HRS) and Hierarchical Equation of Motion (HEOM) for modeling exciton dynamics. A specific attention was paid to the role of structural disorder and dynamical fluctuations in the studied system, where we scan over a broad range of parameters. We systematically obtain that the exciton propagation length should be about 5-10 times shorter than the reported in experiments. This finding suggests that there are may be other physical processes explaining the observations. Several possible mechanism are discussed.

\begin{acknowledgement}

We appreciate discussions with Joel Yuen-Zhou and Dmitrij Rappoport. We are also thankful to Doran Bennett for valuable discussions of the tripole model and non-Markovian exciton diffusion. SKS, CK, and AAG acknowledge the support from the Center for Excitonics, an Energy Frontier Research Center funded by the U.S. Department of Energy under award DE-SC0001088. AAG also thanks the Canadian Institute for Advanced Research for support. UP acknowledges the support of the Adelis Foundation and the Grand Technion Energy Program. MAS and YNP are thankful to the Ministry of Education and Science of the Russian Federation for supporting the research in the framework of the state assignment, award \# 3.2166.2017/4.6. First principles computations were run on Harvard University's Odyssey cluster, supported by the Research Computing Group of the FAS Division of Science.

\end{acknowledgement}

\begin{suppinfo}

\begin{itemize}
  \item SI1: Computed spectra of CBT, NIBT and composed CBT-(NIBT)$_3$ pigments.
  \item SI2: Split-exciton model.
\end{itemize}

\end{suppinfo}

\bibliography{H_agg}

\end{document}